%% file: main.tex
\documentclass[a4paper,twoside]{article}

\usepackage{epsfig}
\usepackage{subcaption}
\usepackage{calc}
\usepackage{amssymb}
\usepackage{amstext}
\usepackage{amsmath}
\usepackage{amsthm}
\usepackage{multicol}
\usepackage{pslatex}
\usepackage{apalike}
\usepackage{algorithm2e}
\usepackage{booktabs}
\usepackage[bottom]{footmisc}

\usepackage{xcolor}

\usepackage{hyphenat}

\usepackage{SCITEPRESS}   

\begin{document}

\title{User Story Tutor (UST) to Support Agile Software Developers}

\author{\authorname{
Giseldo da Silva Neo\sup{1}\orcidAuthor{0000-0001-5574-9260},
José Antão Beltrão Moura\sup{2}\orcidAuthor{0000-0002-6393-5722}, 
Hyggo Almeida \sup{2}\orcidAuthor{0000-0002-2808-8169}, 
Alana Viana Borges da Silva Neo \sup{2}\orcidAuthor{0009-0000-1910-1598}, 
and Olival de Gusmão Freitas Júnior\sup{3}\orcidAuthor{0000-0003-4418-8386}}
\affiliation{\sup{1}Campus Viçosa, Federal Institute of Alagoas, Viçosa, Brazil}
\affiliation{\sup{2}Center for Electrical and Computer Engineering, Federal University of Campina Grande, Campina Grande, Brazil}
\affiliation{\sup{3}Institute of Computing, Federal University of Alagoas, Maceió, Brazil}
\email{giseldo.neo@ifal.edu.br, \{antao,almeida\}@computacao.ufcg.edu.br, alana.neo@copin.ufcg.edu.br, olival@ic.ufal.br}}


\abstract{User Stories record what must be built in projects that use agile practices. User Stories serve both to estimate effort, generally measured in Story Points, and to plan what should be done in a Sprint. Therefore, it is essential to train software engineers on how to create simple, easily readable, and comprehensive User Stories. For that reason, we designed, implemented, applied, and evaluated a web application called User Story Tutor (UST). UST checks the description of a given User Story for readability, and if needed, recommends appropriate practices for improvement. UST also estimates a User Story effort in Story Points using Machine Learning techniques. As such UST may support the continuing education of agile development teams when writing and reviewing User Stories. UST’s ease of use was evaluated by 40 agile practitioners according to the Technology Acceptance Model (TAM) and AttrakDiff. The TAM evaluation averages were good in almost all considered variables. Application of the AttrakDiff evaluation framework produced similar good results. Apparently, UST can be used with good reliability. Applying UST to assist in the construction of User Stories is a viable technique that, at the very least, can be used by agile developments to complement and enhance current User Story creation.}

\onecolumn \maketitle \normalsize \setcounter{footnote}{0} \vfill

\section{INTRODUCTION}

Every year, a report that consolidates data from technology projects from different companies in several countries is published on the internet by the Standish Group International Organization. The report is called Chaos Report \cite{Standish2016}. This report indicates that less than 1/3 of the surveyed software technology projects were completely successful. Most of the projects did not reach completion within the planned time and cost budget estimated.

Software development is a complex process that involves many variables and is prone to several failures \cite{sommerville2011}. Much of this failure is related to the specification of what should be done and other factors. To minimize this problem some methodologies and frameworks, like Agile methods, can provide a conceptual structure for conducting this software engineering projects. 

Agile can be understood as a set of behaviors, processes, practices, and tools used to create products and subsequently make them available to end users \cite{Cohn2005}. One of the best-known representatives of Agile methods is SCRUM \cite{Sutherland2014}. It focuses primarily on the aspect of what must be done.  In SCRUM requirements must be specified at an adequate level of clarity, neither complex nor too rigid. An important part of this method is writing, interpreting, and implementing what is called a User Story. 

A User Story is a short and simple sentence about a feature (written from the perspective of the user who wants it) and is used to define the scope of a software project \cite{cohn2004user}. It is a requirements analysis technique that captures the ``who'', ``what'' and ``why'' concisely and simply, usually limited in detail, so that it can be written by hand on a small note card of paper. These User Stories are generally stored in software that manages all the project life-cycle \cite{jadhav2023systematic}. By analyzing these raw data, recorded in these tools, we can extract information for various software engineering research \cite{Tawosi2022}. 

However, writing a good User Story can be difficult. The User Story can be very shallow and not present adequate detail to understand the expected final result, or, conversely, it can be too comprehensive. For example: A stakeholder may confuse the level of detail of a User Story and write the scope of an entire module or system, which is not appropriate. Also, the quality of User Stories can have a big impact when the agile team makes estimates. The writing of good user stories, which can be used to estimate effort, appears as one of the 5 most important agile problems informed by 119 developers \cite{Andrade2021}.

Improving the creation of the User Story is crucial for better planning and consequently the success of the project. Therefore, the objective of this study is to assist the User Story creation process, recommending improvements, using natural language processing in an intelligent learning environment, i.e., a pedagogical agent that assists in the learning process, characterizing itself as a tutor of content or more adapted strategies. The hypothesis is that this environment can help agile teams build better user stories. An expected contribution is to help development teams that use agile practices to build better user stories.

The tool User Story Tutor (aka UST), proposed in this article, receives as input a User Story text in English and presents personalized recommendations for improving it, with the support of a large language model (LLM). LLMs are very large deep learning models that are pre-trained on vast amounts of data. The tool also presents a prediction in Story Points, generated by a machine learning algorithm trained with data from other projects. The user is also presented with the User Story's readability indexes, as they can be used to represent an indicator of text clarity.

We use the Design Science Research search methodology with 3 phases: problem identification, solution design, and evaluation to build the proposed tool \cite{wieringa2014design}. For evaluation, a survey was carried out with the support of the Technology Acceptance Model (TAM) framework \cite{davis1989user} and the AttrakDiff evaluation framework \cite{hassenzahl2003attrakdiff}. 40 agile practitioners who did not engage in UST's development, evaluated the proposal solution and responded to the Survey. The TAM and AttrakDiff evaluation results indicate that UST meets the established objectives, with good acceptance from participants. 


The remainder of the paper is organized as follows. Section ~\ref{sec:background} summarizes the main concepts that facilitate the understanding of subsequent sections. Section~\ref{sec:methodology} discusses the methodology and methodological artifacts adopted in UST's R\&D. Section 5 brings more technical details of UST. Section~\ref{sec:proposal} addresses UST's evaluation results. Section~\ref{sec:related_work} is devoted to related work. Section~\ref{sec:threats_to_validity} explores threats to the validity of our investigation and its results. Lastly, Section~\ref{sec:final_consideration} offers the final considerations.

\section{BACKGROUND}\label{sec:background}

\subsection{User Stories}

User Stories are a central piece in the development of requirements for teams that use agile development. A characteristic of agile methods is their focus on fast, value-added deliveries in short periods, dealing with changes as quickly as possible \cite{Dyba2008}. This approach has shown results and has been used in project management \cite{PMI2017}, both in industry \cite{Trimble2016,Rigby2018} and in government \cite{Mergel2016}.

SCRUM is an agile method based on fixed time cycles called sprints, where teams work to achieve well-defined objectives; these objectives are represented in the Product Backlog, a list of things to do that is constantly updated and re-prioritized \cite{Sutherland2014}. Software requirements are usually stored in User Stories. These artifacts describe the activities that the development team will estimate and build and are written in natural language. 

Most teams that use User Stories, or even the agile method, also use software tools to manage the project and mainly to keep a record of their user stories \cite{jadhav2023systematic}. By analyzing the data recorded by these tools we can extract information for various software engineering research, including research on how to improve these same User Stories \cite{jimenez2023usqa}.

GitLab is one of these management tools used by agile teams to record User Stories \cite{Choudhury2020}. It allows software engineers to automate many actions during the development cycle, including recording and changing User Stories \cite{dimitrijevic2015comparative}. In GitLab, the User Story is registered as an Issue, and for each User Story various pieces of information are stored, such as the title, the task description, and its estimate in Story Points.

This data stored in these management tools can support decision-making in various software engineering scenarios, such as: assigning User Stories \cite{Mani2019}, improving the description of User Stories \cite{Chaparro2017}, iteration planning \cite{Choetkiertikul2018}, sentiment analysis of developers who write User Stories \cite{Ortu2015,Ortu2016,Valdez2020}, effort estimation of User Stories \cite{Porru2016,Soares2018,Dragicevic2017,Choetkiertikul2019,Tawosi2022Deep}, and prioritization of User Stories \cite{Gavidia-Calderon2021,Huang2021,Umer2020}.

  

\subsection{Recommendation Systems}

Recommendation systems are software tools and techniques that provide suggestions for items that are more likely to arouse the interest of a particular user \cite{ricci2010introduction}. Recommendations are important for the learning process by allowing teachers and students to find content more appropriately, according to their profile and needs. 

The evolution of Recommendation Systems is moving towards a set of hybrid techniques, which combine two or more different Recommendation techniques, to resolve the limitations and obtain the advantages of each of them \cite{bobadilla2013recommender}. A hybrid Recommendation system is a term used to describe any Recommendation system that combines several Recommendation techniques to produce an output. Burke (2007) cites the following types: Collaborative, Content-based, demographic, and knowledge-based \cite{burke2007hybrid}.

Recommendation systems have gained a lot of popularity in the educational field, generating various types of recommendations for students, teachers, and schools. 
They can reduce student information overload by recommending the “right” information at the right time and in the right format of interest to the student \cite{odilinye2020personalized}.



More recently, LLM-type models (for example, ChatGPT from OpenAI) can be also used for recommendations. OpenAI is the company behind an LLM called ChatGPT. They offer this large-scale multimodal model to be used for third parties in situations where reliability is not critical \cite{OpenAI2023}. 

Despite some problems with the LLM recommendation approach: responses from an LLM are not completely reliable, have a limited context window, and do not learn \cite{OpenAI2023}. They can be useful in specific domains. They are suitable in situations where there are well-defined intentions, for example, opening a bank account, scheduling an air ticket, or improving a User Story, given their controlled structure and predictable output. In this study, OpenAI will be used as a synonym for OpenAI API, or even ChatGPT, and in the context of this paper, a recommendation is a text to improve the User Story, which the system suggests to the user.


\subsection{Readability Indexes}

There are some readability indexes commonly used in the literature to predict the reading ease of the text. They are used to determine the readability of an English passage and they are already used in fake news and opinion spam detection. This section describes the 4 most common of them. 

Readability indexes can be interpreted as a numerical indicator of how much easier it is for other people to read writing text \cite{DuBay2004}.  To extract this numerical information some of the algorithms use the count of words, characters, sentences, syllables, and a list of complex words in their formula. 

Readability indexes have been used by educators since 1920. In 1980 there were already 200 known calculation formulas \cite{DuBay2004}. They have already been criticized by researchers, who point out their limitations \cite{koenke1971another}. However, empirical experiments confirmed the relationship between these indexes and the readability of the text \cite{bogert1985defense}.

Gunning’s Fog Index is the most frequently used and studied index and has been extensively used to analyze text \cite{bogert1985defense}. It is a numerical number assigned to a given text that uses words, sentences, and a list of complex words in their formula. The higher the value, the more complex the text.  It was created by Robert Gunner in 1954. The Greater the percentage of complex words, the harder the text is to read. The Higher the index, the lesser the readability. His algorithm and method of calculation are well documented \cite{gross1985fogindex}. It can be computed by adding the average sentence length and the percent of complex words (words of three or more syllables) and multiplying that sum by 0.4. Like in the formula presented in Equation~\ref{eq:equacao_fog_index}.

\begin{equation}
   0.4 \cdot \left[\left(\frac{words}{\text{sentences}}\right)+100 \cdot \left(\frac{words~complex}{words}\right)\right]
   \label{eq:equacao_fog_index}
\end{equation}

Another index of text readability is Flesch Reading Ease, according to \cite{textstat2023}. The higher the value, the more difficult it is to read the text. Its maximum value is 121.22. There is no minimum value, negative scores are also valid. Equation~\ref{eq:fre} presents the calculation of the Flesch Reading Ease. It is one of the oldest and most widely used tests and is only dependent on two factors: The Greater the average sentence length, the harder the text is to read. The greater the average number of syllables in a word, the harder the text is to read. The higher the score, the greater the readability.

\begin{equation}
206.835
-1.015 \cdot \left(\frac{words}{sentences}\right) 
-84.6 \cdot \left(\frac{syllables}{words}\right)
\label{eq:fre}
\end{equation}

Coleman Liau Index is another complexity index \cite{textstat2023}, but this time using another Equation~\ref{eq:cli}. Where L is the average number of letters per 100 words, and S is the average number of sentences per 100 words.

\begin{equation}
CLI= 0.0588 \cdot L - 0.296 \cdot S - 15.8 \\
\label{eq:cli}
\end{equation}

Finally, the Automated Readability Index is calculated from the following Equation~\ref{eq:ari} \cite{textstat2023}.

\begin{equation}
    4.71 \cdot
    \left(
    \frac{{characters}}{{words}}
    \right)
    +0.5 \cdot 
    \left(
    \frac{{words}}{{sentences}}
    \right)
    -21.43    
    \label{eq:ari}
\end{equation}




\section{METHODOLOGY}\label{sec:methodology}

The research methodology used in this research was Design Science Research \cite{wieringa2014design}. We designed a web application called User Story Tutor (UST) that uses Natural Language Processing, Readability Indexes and Machine Learning Prediction as a proof of concept to improve User Story writing. We used a survey, supported by a questionnaire in Google Forms, to evaluate UST. The development was carried out in the following stages:

\begin{itemize}
    \item Literature review;
    \item Design of a predictive model that predicts the number of Story Points from the User Story using Machine Learning;
    \item Definition of the basic text readability indexes that can be extracted from User Stories;
    \item Design of a recommendation module via querying the OpenAI API;
    \item Implementation all 3 modules as a web application;
    \item Internal evaluation using many User Stories from real projects;
    \item Evaluation with participants in a Survey;
    \item Qualitative and quantitative analysis of the Survey results.
\end{itemize}

To evaluate UST, we carried out a survey based on the Technology Acceptance Model (TAM) framework \cite{davis1989user} and AttrakDiff evaluation framework \cite{hassenzahl2003attrakdiff}. TAM is an information systems theory that models how users accept and use technology - please see Figure 1. For the TAM statistical test, Cronbach's alpha was used. The AttrakDiff \cite{hassenzahl2003attrakdiff} test presents quality factors (hedonic and pragmatic) that can help to better evaluate the proposal, complementing the TAM framework. The Survey collected participants' perceptions and suggestions regarding UST, whose objective is to assist agile practitioners in building better User Stories. The questions used in the survey are presented in Section~\ref{sec:RESULTS}.

\begin{figure*}
    \centering
    \includegraphics[width=0.7\linewidth]{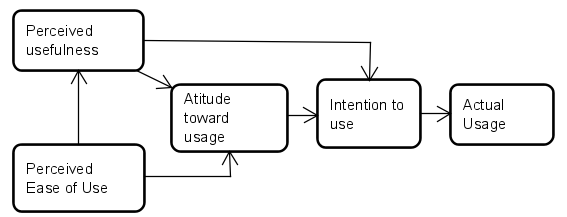}
    \caption{TAM. Adapted from \cite{ALOMARY2015}}
    \label{fig:enter-label}
\end{figure*}

UST's architecture was divided into 3 modules (Recommender, Estimator, and Readability) and for each module, certain procedures were selected and executed. These selected general procedures will be detailed in the following paragraphs.

The Recommender module is responsible for recommending improvements to User Stories by returning text in natural language. The LLM model used (made available by OpenAI) is already trained with a large amount of text, extracted from the entire internet \cite{torrentcopilots2023}. However, customization of the prompt is necessary to better personalize the return. 


To customize the response from the OpenAI LLM model, we send a prompt to configure the return according to the needs of the recommendation system. Following
OpenAI’s guidelines for building effective prompts, 3 main recommended techniques were used in the prompt design.


\begin{enumerate}
    
    \item Clarity in instruction
    
    We seek a clear and precise prompt to not generate doubts when returning the recommendation. The probability of a good return recommendation depends on the objectivity of the hidden prompt sent along with the recommendation. 
    
    

   \item Split complex tasks into simpler tasks
    
    Intending to limit the task, we sent (a prompt) text to limit the set of return possibilities, as complex tasks generally have a higher error rate than simpler task requests;

   

    

    \item Test changes
    
    Several interactions were necessary to create the prompt used in search of improvements and following the good practices recommended by OpenAI itself.
    
\end{enumerate}

The Estimator module uses a supervised learning algorithm, selected according to its best prediction capacity to predict the Story Points from the informed User Story text. To elaborate the predictor module we followed the techniques: data collection; data exploration; data preparation; creation, training, and validation; adjustment of hyperparameters, and implementation of the model. 

The User Story Estimator predictor was trained with other User Stories from other agile projects. This dataset was collected from real open-source projects extracted from an open-source repository. More details of the dataset are presented in Section~\ref{dataset}. The metric used to select the best algorithm and for hyperparameters adjusting was the Mean Absolute Error with cross-validation. In the end, The predictor model was trained with all the data and made available for the proposed application. 

Finally, the module related to Readability presents 4 text readability indexes: Gunning Fog, Automated readability index, Coleman Liau Index, and Flesch Reading Ease (please refer to Section~\ref{sec:background}). But it is important to highlight that for this proposal, to facilitate the interpretation of the readability index in general, a variable called Final Result was also created, which is the arithmetic mean of the 4 selected Indexes. 

\section{THE USER STORY TUTOR}\label{sec:proposal}

This section presents an overview of the technologies used to build the User Story Tutor (aka UST), its high-level architecture, the dataset used, and its application interface.

Our idea is that UST supports agile teams during the construction of User Stories and assists the development process during the User Story preparation phase and in estimating task effort. UST consists basically of a web application that can be accessed using a browser on mobile devices, PCs, and notebooks. UST uses an LLM provided by OpenIA to recommend improvements and present readability indexes. The main parts of the tool are discussed below.

\begin{figure*}[!htb]
    \centering
    \includegraphics[width=\linewidth]{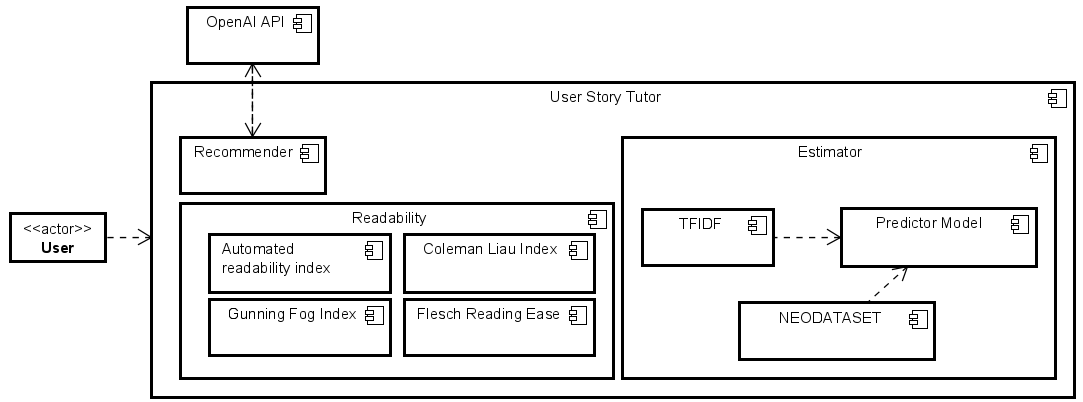}
    \caption{UST Architecture}
    \label{fig:architecture}
\end{figure*}

The application was designed around 3 modules with well-defined functions. The Recommender module responds to User Story recommender requests. It is responsible for maintaining the prompt and combining it with the text of the new User Story, querying OpenAI via API, and preparing the return for presentation to the user. The module that performs User Story estimates in Story Points uses a predictive model, already trained with historical data to assist developers in their estimates, acting as a reference for the team responsible for estimating effort. Readability indexes are extracted from text with basic natural language processing techniques. An image of the architecture is presented in Figure~\ref{fig:architecture}.

For coding, we used StreamLit \footnote{https://streamlit.io} - a library for building open-source applications for machine learning and data science. Python \footnote{https://www.python.org/} was used as the language - a programming language that has been increasing its market share, mainly in applications that use machine learning. The Recommender Module performs a query to OpenAI. The scikit-learn libraries were also used \footnote{https://scikit-learn.org}. All source-code of the project was available at Github \footnote{https://github.com/giseldo/userstory}. The UST can be tested at StreamLit Cloud \footnote{https://userstoryteach.streamlit.app}. 

\subsection{Dataset}\label{dataset}

We have made a new dataset (aka NEODATASET) available together with UST. This dataset encompasses data from \input{conf/projects} software development projects, with \input{conf/tasks} User Stories taken from GitLab repositories, totaling \input{conf/story_points} Story Points. It is made available on GitHub \footnote{https://github.com/giseldo/neodataset} so that the entire interested community can contribute, similarly to what happens with other datasets.

This dataset was mined during January 2023 and April 2023. The mining process targeted GitLab's top open-source projects. The selected projects employ agile software development methodologies and had the size of their tasks recorded in \textit{Story Points}. To mine information from GitLab, we created an extraction tool implemented in Python that connects to GitLab via API


Only Tasks with the State attribute equal to Closed and that have the \textit{weight} attribute filled in were collected. The \textit{weight} field is used in GitLab to record the effort in Story Points. More information about the projects included in the dataset is also available directly from GitLab.


The projects in the dataset have different characteristics and cover different programming languages, different business domains, and different geographic locations of the team. The main entity of the dataset is the User Story (or Issue), which contains the main information. The dataset has more than \input{conf/attributes} attributes and is stored in JSON and CSV format, given the simplicity of dealing with both formats.


The dataset presented here includes projects which were not used by previous studies. There are already previous studies that extracted data from the Jira management tool to build predictive models \cite{Tawosi2022,Choetkiertikul2019,Porru2016,Scott2018}, but projects extracted from GitLab are rarer.

Just as \cite{Tawosi2022} did, we are sharing all the data collected. Therefore, the most common thing is to share only the data from the dataset considered in the study itself, as done, for example in \cite{Choetkiertikul2019}, and not all the data collected.

The expected contribution is that this data set can assist education and research on agile software development. Although our dataset was initially designed for Story Points and User Story estimation training and research, it also includes information relevant to other software engineering aspects. In addition to providing a possibility to reproduce findings from other studies.

\subsection{User Interface}

\begin{figure}[!htb]
    \centering
    \includegraphics[width=\linewidth]{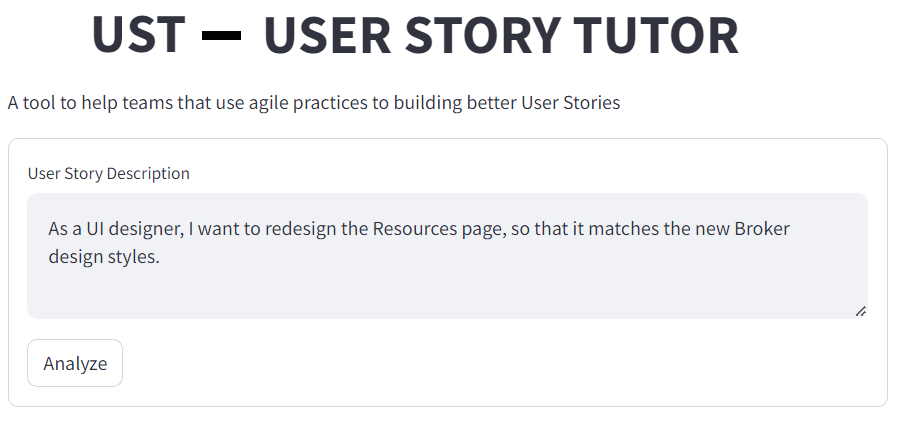}
    \caption{Home Screen}
    \label{fig:tela_principal}
\end{figure}

The first screen of the UST (Figure~\ref{fig:tela_principal}) is where a developer from the agile team informs their User Story. Any User Story from any real project can be used. Then, after the developer enters the description of the User Story in text format, he clicks on the ``Analyze'' button. The UST then initializes the necessary threads that trigger the responses of the existing modules. The language of the UST interface was English.

\begin{figure}[!htb]
    \centering
    \includegraphics[width=\linewidth]{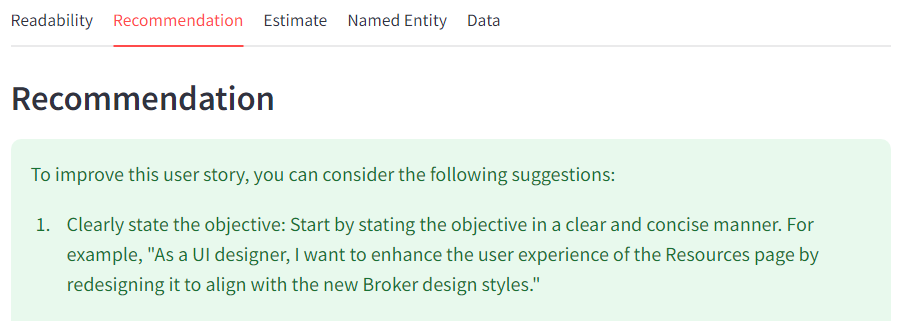}
    \caption{Recommendation Module Example}
    \label{fig:tela_recomendacao}
\end{figure}

The recommender module is a module that returns suggestions from the developer Story description text entered by the user (Figure~\ref{fig:tela_recomendacao}). The model provided by OpenAI was used, specifically the gpt-3.5-turbo. The parameters used by the recommender module are sent via a hidden prompt, and they are presented in Table~\ref{tab:prompt}. They followed a refinement process similar to creating a search key in a systematic literature review, being revised and adapted until we reached the final version presented with the support of the produced data set in Section~\ref{dataset}.

\begin{table}[!htb]
    \centering
    \caption{OpenAI personalized prompt}
    \begin{tabular}{cp{5.4cm}}
         \toprule
         \textbf{role} & \textbf{content} \\
         \midrule
         system & You are a scrum master, skilled in creating better User Stories for agile software projects \\
         user & How can I improve this user story: + \textit{User Story text informed by user in english} \\
         \bottomrule
    \end{tabular}
    \label{tab:prompt}
\end{table}

For the Recommender module, a valid input from the user would be this imagined User Story: \textit{``As a UI designer, I want to redesign the Resources page, so that it matches the new Broker design styles.''} The Recommendation text that UST would return to the user is: \textit{`` To improve this user story, you can consider adding more details and making the objective clearer. Here's an improved version of the user story: As a UI designer, I want to redesign the Resources page to improve the user experience and align it with the new Broker design styles, enhancing the overall consistency and visual appeal of the application. Additionally, you can further refine the user story by specifying the specific changes or improvements you plan to make to the Resources page''}.

\begin{figure}[!htb]
    \centering
    \includegraphics[width=\linewidth]{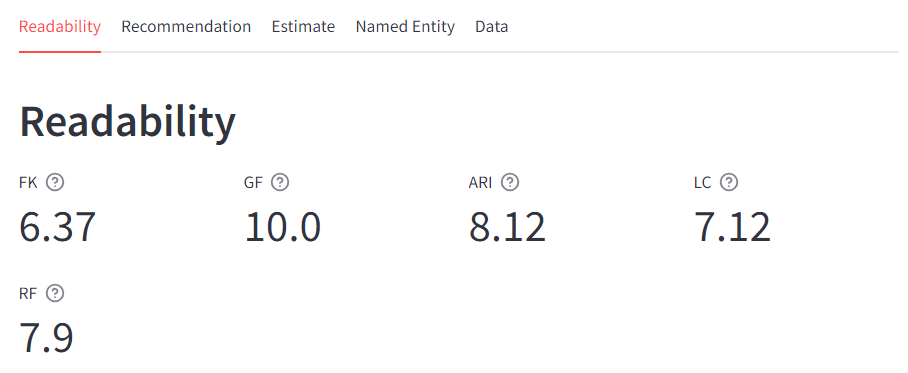}
    \caption{Readability Indexes Module}
    \label{fig:tela_principal_readability}
\end{figure}

For the User Story readability module (Figure~\ref{fig:tela_principal_readability}), the readability indexes of the text are extracted using the textdescriptive library (available in Python) \footnote{https://pypi.org/project/textdescriptives} and presented on the screen with the ``metrics'' component from StreamLit in the first tab. The purpose of the readability module is to allow the creator of the User Story to see some quantitative measure of how easy the text of their User Story is to read.

\begin{figure}[!htb]
    \centering
    \includegraphics[width=\linewidth]{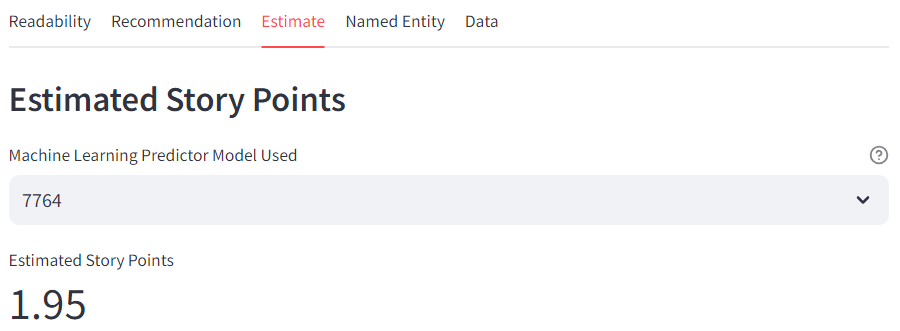}
    \caption{Estimation Module}
    \label{fig:tela_principal_estimate}
\end{figure}

Finally, the Effort Estimation module (Figure~\ref{fig:tela_principal_estimate}) performs an effort estimate in Story Points based on the User Story description. The predictive model and vectorizer used are loaded with the Joblib library. The selected algorithm was SVM. The User Story text is transformed into a bag-of-words using the TFIDF technique. In production, both the vectorizer and the model are loaded into memory for prediction. After the User Story text is transformed into a matrix with the vectorizer, it is passed on to the model predictor, which returns the estimate in Story Points. The loaded Model was previously trained with data from NEODATASET.

\section{EVALUATION}\label{sec:RESULTS}

This section presents a qualitative and quantitative evaluation of the tool with the support of the TAM framework and AttrakDiff and discusses the results. The survey was conducted in December 2023 with an online questionnaire in Google Forms. The questionnaire was first examined for comprehensiveness, quality and adequacy to the investigation at hand by a panel of 6 experts  who had amongst them 7 years of experience in agile development. The questionnaire used a 5 level Likert scale to gauge the respondent's agreement (from none or level 1), passing through little (2), neutral (3), somewhat (4) to full (level 5) with statements made concerning UST. The experts' comments and suggestions led to the adjustment of the questionnaire which was then applied to a sample of respondents. 


\subsection{Sample Characterization}

Our sample of survey respondents is made up of 40 Brazilian participants. 70\% of those who responded to the survey had worked directly with agile methodologies. More than half of these (56\%) had worked in a software factory and had already worked as a member of a software development team. Of these, 20\% have been Scrum Masters and 10\% have been Product Owners. The other participants had generally participated in academic activities related to software engineering. On average, our sample was made up of professionals with 3 years of experience in agile methodology.

\subsection{TAM}

Each of the 4 constructs of the Technology Assessment Model is analyzed below: perception of usability, perception of ease of use, external variables, and attitude. 

The perception of usability is the level at which a person believes that using UST improves the performance of their tasks. To analyze the perception of usability, (Table~\ref{tab:perceptionofease}) the mean, median, and standard deviation of the Likert scale responses were analyzed. If the Mean or Median is above the threshold which we chose to be "3" (neutral) in our experiments, this possibly indicates that the participants have a positive attitude towards the perception of using the tool \cite{dantas2019effort}. 

Evaluating the responses, it is possible to infer that participants generally have a positive attitude toward the perceived usability of the tool (Table~\ref{tab:perceptionofease}).

\begin{table}[!htb]
\caption{Perception of usability}
\label{tab:perceptionofease}
\centering
\begin{tabular}{p{4cm}p{0.6cm}p{0.5cm}p{0.5cm}}
    \toprule
    \textbf{Definition} & \textbf{Mean} & \textbf{Med} & \textbf{SD} \\
    \midrule
    V1 Using the tool is useful to improve my User Stories & 4.45 & 5.00 & 0.80 \\
    \addlinespace
    V2 I learned how to build better User Stories after using the tool & 4.07 & 4.00 & 1.10 \\
    \bottomrule
\end{tabular}
\end{table}

The perception of ease of use is the Level at which the person presents their perception of the tool in terms of ease of learning and operation. Table~\ref{tab:percepcaodefacilidade} describes the mean, median, and standard deviation of the responses related to perceived ease of use. All averages are above the adopted threshold, therefore also regarding good perception and ease of use of UST. A standard deviation above one indicates a high dispersion in responses. 

\begin{table}[!htb]
\caption{Perceived ease of use}
\label{tab:percepcaodefacilidade}
\centering
\begin{tabular}{p{4cm}p{0.6cm}p{0.5cm}p{0.5cm}}
    \toprule
    \textbf{Definition} & \textbf{Mean} & \textbf{Med} & \textbf{SD} \\
    \midrule
    V3 Learning to use the tool was easy for me & 3.70 & 4.00 & 1.28 \\
    \addlinespace
    V4 Searching for information in this tool was simple & 3.72 & 4.00 & 1.09 \\
    \addlinespace
    V5 Accessing the tool is simple & 4.07 & 4.00 & 1.14 \\		
    \bottomrule
\end{tabular}
\end{table}

An analysis of external variables, which provides a better understanding of what influences perceived utility and ease of use, is presented in Table~\ref{tab:externalvariables}. A median above 4 is a good indicator that the external characteristics were well accepted by users. 

\begin{table}[!htb]
\caption{External variables}
\label{tab:externalvariables}
\centering
\begin{tabular}{p{4cm}p{0.6cm}p{0.5cm}p{0.5cm}}
    \toprule
    \textbf{Definition} & \textbf{Mean} & \textbf{Med} & \textbf{SD} \\
    \midrule
    V6 The application's navigation attributes - menu, icons, links, and buttons - are clear and easy to find & 4.02 & 4.00 & 1.17 \\
    \addlinespace
    V7 The tool has a good interface & 3.9 & 4.00 & 1.20 \\
    \bottomrule
\end{tabular}
\end{table}

The data characterized as Attitude, which is the Intention of the individual to use the tool, are presented in Table~\ref{tab:atitude}. In the same way, as with the other constructs, we have a mean above the threshold. 

\begin{table}[!htb]
\caption{Attitude}
\label{tab:atitude}
\centering
\begin{tabular}{p{4cm}p{0.6cm}p{0.5cm}p{0.5cm}}
    \toprule
    \textbf{Definition} & \textbf{Mean} & \textbf{Med} & \textbf{SD} \\
    \midrule
    V8 I believe it is better to use the tool to help create the user story than not to use it. & 4.25 & 4.00 & 0.88 \\
    \addlinespace
    V9 I intend to use the tool to create better user stories and to plan my tasks better & 3.85 & 4.00 & 1.15 \\
    \bottomrule
\end{tabular}
\end{table}

For statistical confirmation, Cronbach's \cite{gliem2003calculating} test was used for the Likert scale questionnaire, the same technique used by \cite{dantas2019effort}. Cronbach’s Alpha is a way to measure the internal consistency of a questionnaire or survey. Cronbach’s Alpha ranges between 0 and 1, with higher values indicating that the survey or questionnaire is more reliable. An interpretation of Cronbach's alpha is presented in Table~\ref{tab:cronbach_alpha}. 

\begin{table}[!htb]
    \centering
    \caption{Internal consistency from the Survey. Adapted from \cite{zach2023statology}}
    \begin{tabular}{ll}
         \toprule
         \textbf{Cronbach Alpha} & \textbf{Internal consistency} \\
         \midrule
         $ 0.9 \leq \alpha$	& Excellent \\
         $ 0.8 \leq \alpha < 0.9 $ &	Good \\
         $ 0.7 \leq \alpha < 0.8 $ &	Acceptable\\
         $ 0.6 \leq \alpha < 0.7 $ &	Questionable\\
         $ 0.5 \leq \alpha < 0.6 $ &	Poor\\
         $ \alpha < 0.5 $ & Unacceptable \\
        \bottomrule
    \end{tabular}
    \label{tab:cronbach_alpha}
\end{table}

A limit adopted in this research is the Cronbach alpha index greater than 0.7 for the variables analyzed and with confidence in the 95\% range. From the reported values, as shown in Table~\ref{tab:descriptive}, we understand that almost all constructs analyzed are above the established limit. This leads to the conclusion that the internal consistency of this survey is acceptable.

\begin{table}[!htb]
\caption{Cronbach of TAM constructs}
\label{tab:descriptive}
\centering
\begin{tabular}{p{2.8cm}cc}
    \toprule
    \textbf{constructs} & \textbf{Cronbach} & \textbf{IC} \\
    \midrule
    Usability & 0.81 & [0.64 0.90] \\
    Ease of use & 0.92 & [0.87 0.95] \\
    External variables & 0.87 & [0.77 0.93] \\
    Attitude & 0.73 & [0.49 0.85] \\
    \bottomrule
\end{tabular}
\end{table}

\subsection{AttrakDiff}

\begin{figure}[!htb]
    \centering
    \includegraphics[width=\linewidth]{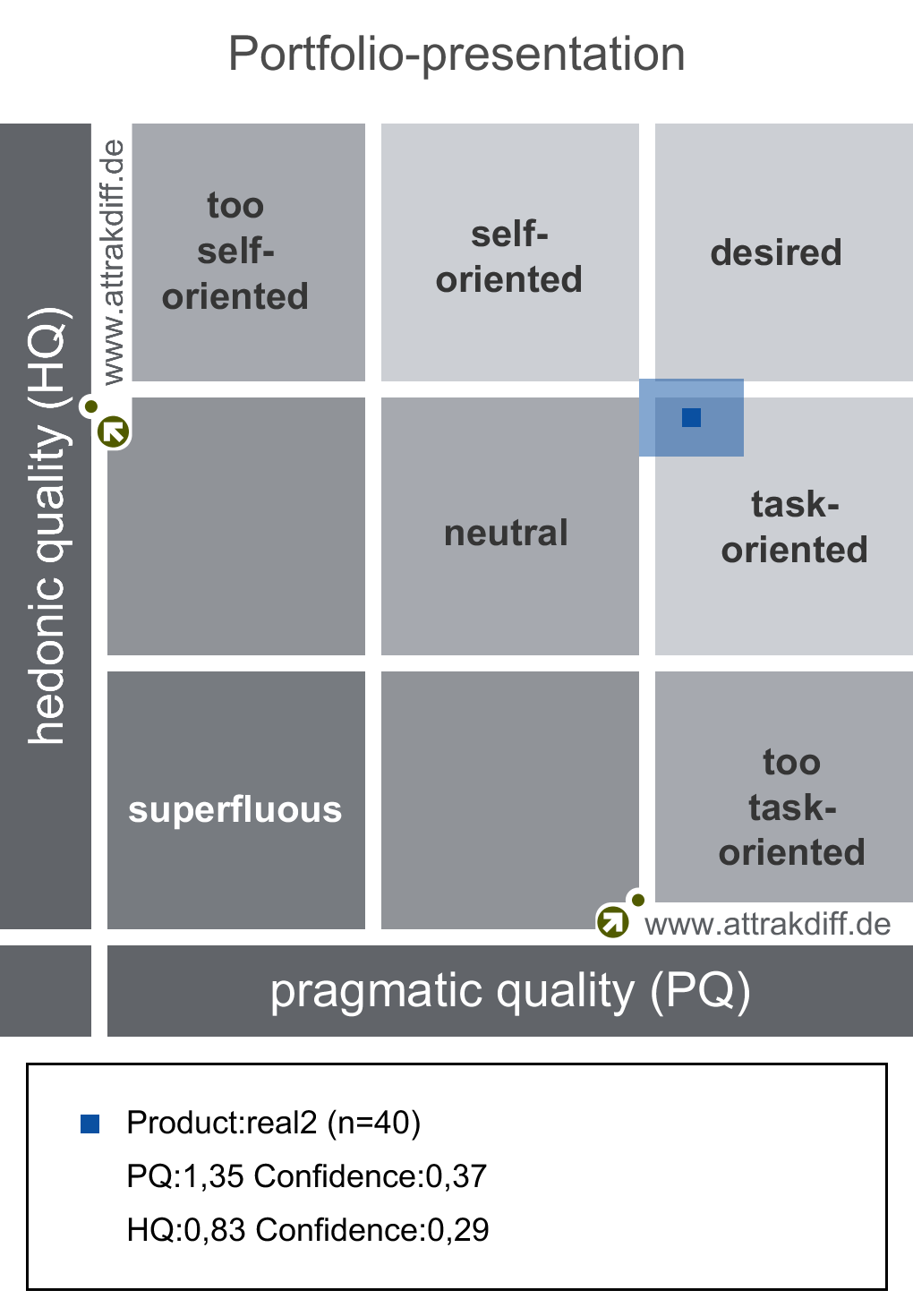}
    \caption{Portfolio of results}
    \label{fig:Portfolio_of_results}
\end{figure}

In Figure~\ref{fig:Portfolio_of_results} we present the portfolio of results of the AttrakDiff test \cite{hassenzahl2003attrakdiff}. This test presents factors that can help better evaluate the proposal, complementing what the TAM framework presents. The AttrakDiff short test type was used, which presents 10 questions to users and infers the metrics reported below.


In Figure~\ref{fig:Portfolio_of_results} the vertical axis of the portfolio view displays the hedonic quality (bottom = low extent). The horizontal axis shows the pragmatic quality (left = low extent). Depending on the dimensions values the product will lie in one or more character regions. The bigger the confidence rectangle, the less sure one can be about which region it belongs. A small confidence rectangle is an advantage because it means that the investigation results are more reliable and less coincidental. The confidence rectangle shows if the users are at one in their evaluation of the product. The bigger the confidence rectangle, the more variable the evaluation ratings \cite{hassenzahl2003attrakdiff}. So, the answers point to a small trust rectangle in the upper right quadrant, as a task-oriented tool.

\begin{figure}[!htb]
    \centering
    \includegraphics[width=\linewidth]{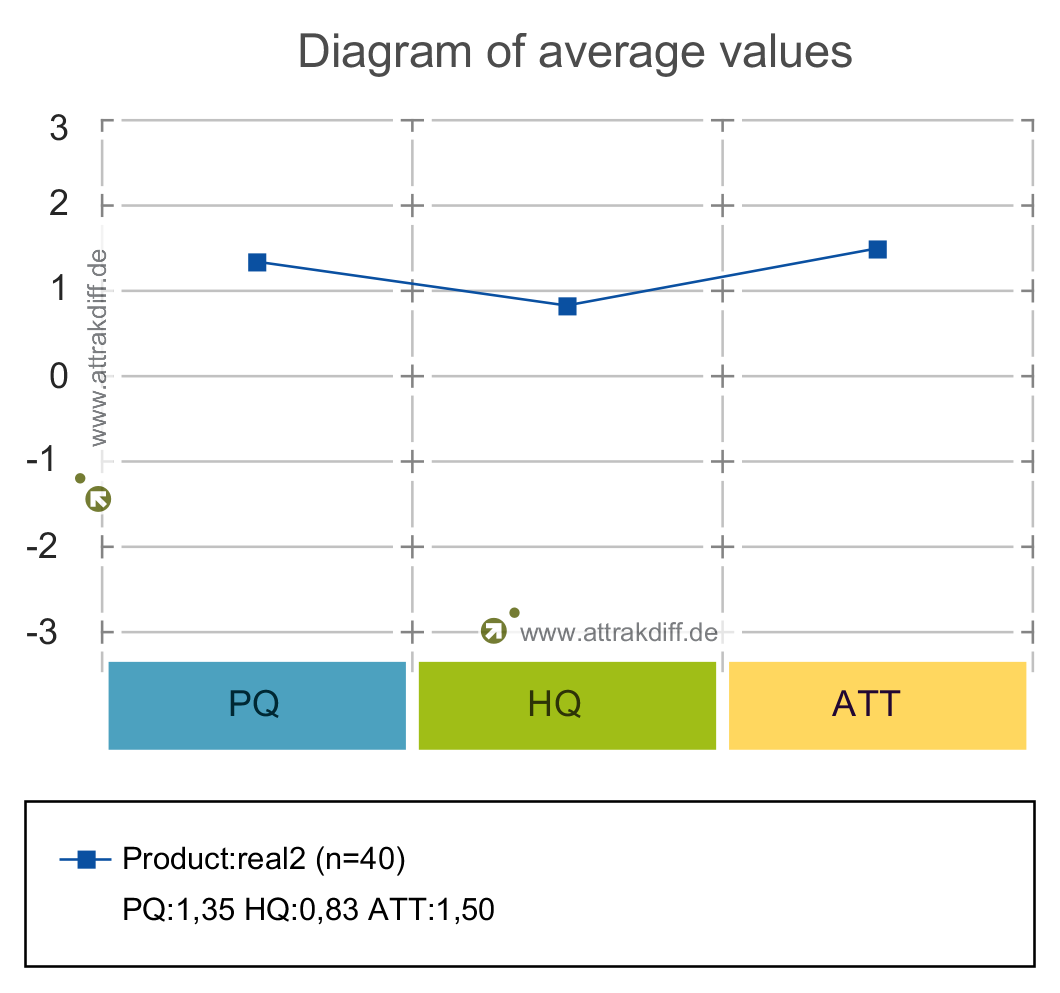}
    \caption{Diagram of average values}
    \label{fig:Diagram_of_average_values}
\end{figure}

In Figure~\ref{fig:Diagram_of_average_values} we present the diagram of average values. The average values of the AttrakDiff dimensions for the evaluated product are plotted on the diagram. In this presentation, hedonic quality distinguishes between the aspects of stimulation and identity. Furthermore, the rating of attractiveness is presented \cite{hassenzahl2003attrakdiff}. 

\begin{figure}
    \centering
    \includegraphics[width=\linewidth]{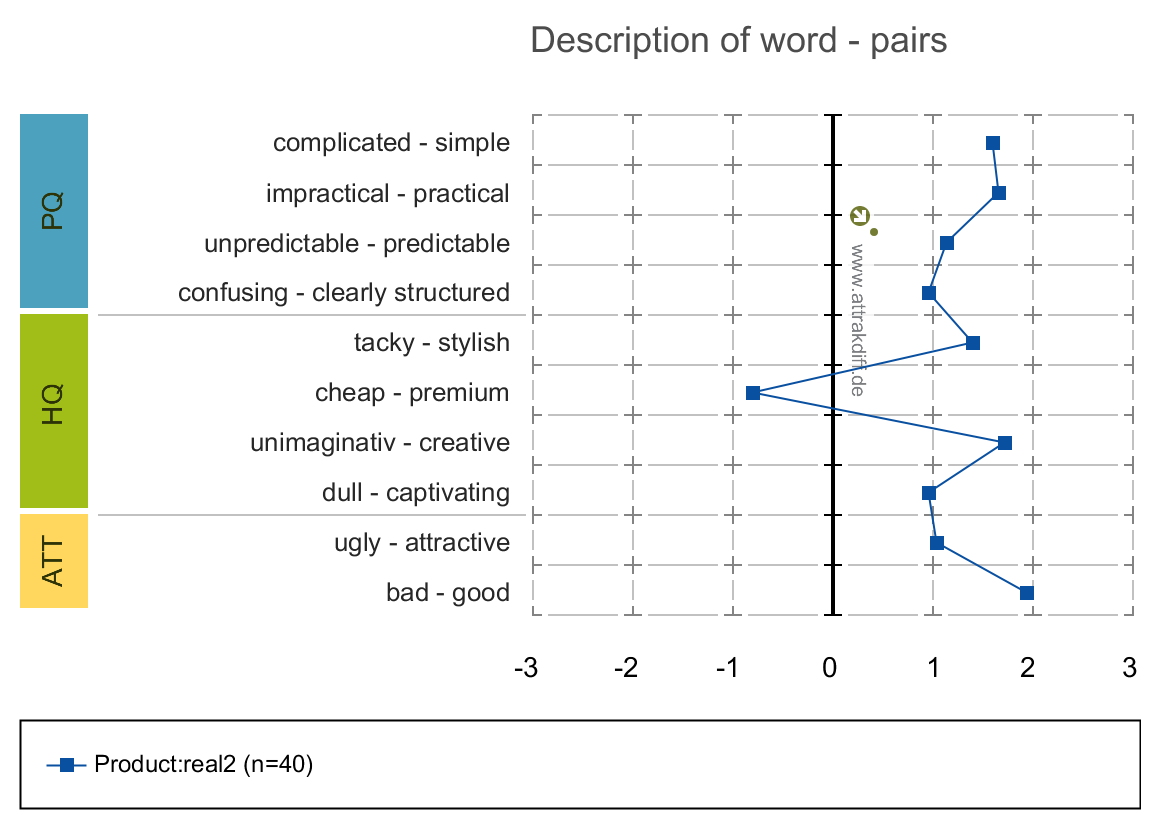}
    \caption{Description of word-pairs}
    \label{fig:Description_of_word-pairs}
\end{figure}

In Figure~\ref{fig:Description_of_word-pairs} we present the description of word pairs. The mean values of the word pairs are presented here. Of particular interest are the extreme values. These show which characteristics are particularly critical or particularly well resolved \cite{hassenzahl2003attrakdiff}. Better results are placed in the positive quadrant, which can be inferred from the consolidated results in Figure~\ref{fig:Description_of_word-pairs}.  Almost all items evaluated were in the positive quadrant, except the pair (cheap-premium). 

\section{RELATED WORK}\label{sec:related_work}

Improving the quality of user stories is a line of research that is gaining momentum due to advances in artificial intelligence. Generally, the most classic approaches use the transformation to an intermediate model, such as a use case \cite{elallaoui2018automatic}, or some other natural language processing technique, with a presentation of reports that can be interpreted (e.g. AQUSA) \cite{lucassen2016improving}.

Using an intermediate model to represent the User Story adds complexity to the use of the solution, which can be seen as an unwanted feature. Our approach does not use an intermediate model. Our approach makes use of text readability indexes, a technique already widely used for text analysis in other areas (e.g. economics, literature), and the use of OpenAI's LLM with personalized recommendations. In addition, the estimator module uses machine learning with natural language processing techniques.

The proposal USQA uses natural language processing techniques to analyze usefulness, completeness, and polysemes in the user stories creation \cite{jimenez2023usqa}. Our proposal brings additional techniques, such as recommendation and readability of the User Story that can help even more. Table 8 compares UST to the AQUSA and USQA proposals and it illustrates UST's contribution as compared to that of existing related work to User Story writing. 

\begin{table}[!htb]
    \centering
    \caption{Comparison with related work}
    \begin{tabular}{p{2cm}p{2cm}p{2cm}c}
         \toprule
         \textbf{Tool} & \textbf{Intermediate Model} & 
         \textbf{Recommend Report} \\
         \midrule
         UST & No & Yes \\
         \addlinespace
         AQUSA & No & No \\
         \addlinespace
         USQA & No & Yes \\
         \bottomrule
    \end{tabular}
    \label{tab:my_label}
\end{table}


\section{LIMITATIONS AND THREATS}\label{sec:threats_to_validity}

There is some criticism in the literature regarding the numerical interpretation of a Likert scale questionnaire  (For example, in the calculation of the Likert scale average or mean) \cite{favero2017manual}. To minimize this point, we use another framework for analyzing software quality, the AttrakDiff.

The use of an LLM model made available by companies via API (e.g. OpenAI's ChatGPT) ties the UST solution to a corporate company. In future work, we intend to use and validate an open-source LLM model.

Readability indexes must be interpreted with caution, as their formulae use only two variables: complex words and long sentences. Therefore, they are not able to measure the cohesion and coherence of a business User Story, which covers semantic, syntactic, and pragmatic factors.

Estimation in Story Points generally follows the Fibonacci scale. In our proposal, the estimator returns a real number between 0 and 100. This problem was treated as a regression problem and not a classification one. However, we can obtain probably greater interpretability if we use the Fibonacci scale instead of real numbers.

\section{CONCLUSIONS AND ONGOING WORK}\label{sec:final_consideration}

This paper presented a proposal and evaluation of a tool for recommending good practices in writing User Stories with LLM, in addition to a User Story estimation module with Machine Learning and presentation of readability indexes for the User Story description. The proposed tool was evaluated by 40 software engineering practitioners. The evaluation was conducted with the TAM and AttrakDiff frameworks. Results indicate that UST meets the established objectives, with good acceptance from its intended users. 

From this investigation, one may conclude that a tool to assist the construction of User Stories is a viable technique that, at the very least, can be used to educate teams on writing better User Stories. In fact, from the evaluation experiment, one may say that UST could help the User Stories by providing feedback to the agile practitioner. 


The paper also presented a dataset with data from projects mined from GitLab that were used to train the predictive model for Story Points. This dataset can be used in other research related to agile software development. Work on named entity recognition to extract entities from the User Story text is ongoing. Independent future work could entail additional validation experiments including integration and evaluation of UST with computer-based education platforms for agile software development methods. 


\bibliographystyle{apalike}

{\small
\bibliography{ref}}

\end{document}

%% file: conf/projects.tex
34

%% file: conf/tasks.tex
40.014

%% file: conf/story_points.tex
163.897

%% file: conf/attributes.tex
70